# Developing Dipole-scheme Heterojunction Photocatalysts

Xu Gao[1,‡], Yanqing Shen[1,‡,*], Jiajia Liu[1], Lingling Lv[1], Min Zhou[1], Zhongxiang Zhou[1], Yuan Ping Feng[2], Lei Shen[3,*]

[1] *School of Physics, Harbin Institute of Technology, Harbin 150001, PR China*
[2] *Department of Physics, National University of Singapore, Singapore 117551, Singapore*
[3] *Department of Mechanical Engineering, National University of Singapore, Singapore 117575, Singapore*

**ABSTRACT:** The high recombination rate of photogenerated carriers is the bottleneck of photocatalysis, severely limiting the photocatalytic efficiency. Here, we develop a dipole-scheme (D-scheme for short) photocatalytic model and materials realization. The D-scheme heterojunction not only can effectively separate electrons and holes by a large polarization field, but also boosts photocatalytic redox reactions with large driving photovoltages and without any carrier loss. By means of first-principles and GW calculations, we propose a D-scheme heterojunction prototype with two real polar materials, PtSeTe/LiGaS$_2$. This D-scheme photocatalyst exhibits a high capability of the photogenerated carrier separation and near-infrared light absorption. Moreover, our calculations of the Gibbs free energy imply a high ability of the hydrogen and oxygen evolution reaction by a large driving force. The proposed D-scheme photocatalytic model is generalized and paves a valuable route of significantly improving the photocatalytic efficiency.

## INTRODUCTION

Photocatalytic water splitting can sustainably convert the inexhaustible solar-energy and abundant water sources into the green and high energy-density hydrogen energy and high-added-value chemicals.[1, 2] The discovery and rational design of photocatalysts is crucial for the photocatalytic water-splitting process because both the light absorption efficiency and photocatalytic redox ability are strongly decided by the catalysts. Unfortunately, a single photocatalyst has a paradox: a narrow band gap is required for a wide range of solar light absorption, while a wide band gap is desired in order to drive the redox ability.[3] Photocatalytic heterojunctions, a photocatalytic system that consists of two different catalysts, have been well-demonstrated experimentally to solve the above dilemma in the single photocatalyst because the specific reduction and oxidation can occur at two different materials. [2, 4-6] The band gap of each material as well as the heterojunction can be even less than the redox energy of 1.23 eV because of the interfacial band alignment and bending. However, the overall photocatalytic efficiency of the heterojunction photocatalysts is still very low due to the rapid recombination of photogenerated electrons and holes – the Achilles heel of water-splitting photocatalysts. So far, several schemes of heterojunction photocatalysts have been proposed and developed to overcome the aforementioned carrier-recombination problem, such as the type-II staggered model and direct Z-scheme. In the type-II heterojunction, the interparticle transfer strategy is proposed to avoid carrier recombination (**Figures 1a-1c**). Under solar light, the photoexcited electrons transfer from the higher conduction band minimum (CBM) of material A to the lower CBM of material B, while the holes move oppositely on the valance band maximum (VBM), resulting the spatial separation of carriers (**Figure 1c**). Consequently, material A and B is the oxidation and reduction catalyst, respectively. However, it is found that this "ideal" charge-transfer route has fundamental problems. In fact, it is hard to obtain such desired charge separation in the described mechanism mainly because of the two following issues. i) The interfacial electric filed (from the lower work-function material A towards the higher work-function material B prevents the electrons (holes) moving along the targeted route (**Figure 1c**), impeding the realization of carrier separation. ii) The proposed improvement of the carrier-separation efficiency is under the sacrifice of the reaction driving forces ($U_e, U_h$) (especially $U_e$ see **Figure 1c**), resulting a low redox ability. Later, direct Z-scheme heterojunctions are proposed for providing large redox driving forces, but they suffer from the same problems of its "Z-scheme family" (see Ref. 7 and references therein).[7] Furthermore, only half of the photoexcited carriers are involved in the Z-scheme photocatalytic reactions, halving the solar-to-fuel efficiency. Therefore, towards highly efficient photocatalytic water splitting, it is important and necessary to develop a novel model with new mechanism with a high carrier-separation efficiency, large redox driving force, and abundant electrons and holes involved in the photocatalytic reactions.

Recently, monolayer or multilayer two-dimensional (2D) polar materials, such as $M_2X_3$ (M = Al, Ga, In; X = O, S, Se, Te), [8-16] IV-VI compounds, [17-19] Li-III-VI compounds, [20, 21] $In_2OX$ (X = S, Se, Te), [22] MXY (M = Mo, W, Cr, Pt; X, Y = O, S, Se, Te; X ≠ Y), [23-33] $MSiGeN_4$ (M = Mo, W), [34] $Sc_2C$, [35] $C_3N_5$, [36] $B_2P_6$, [37] and $CuInP_2S_6$, [38] as well as 2D ferroelectric polar materials, such as $AgBiP_2Se_6$ [39] and $In_2Se_3$,[40] have attracted great attention in the photocatalytic area. Leveraging the intrinsic dipoles in these 2D polar materials, we develop a new photocatalytic model of dipole-scheme (D-scheme for short) heterojunctions (**Figures 1d-1f**), and carry out the material realization by first-principles calculations. Compared with the single photocatalysts and type-II/Z-scheme heterojunction photocatalysts, the D-scheme heterojunction can significantly improve i) the charge-separation efficiency by a large polarization field through the superposition of dipoles of two polar materials (**Figure 1e**), and ii) redox ability by large redox driving forces and maximal utilization of photogenerated carriers (**Figure 1f**). To demonstrate the feasibility and reliability of the D-scheme model, we rationally design a heterojunction with two reported 2D polar materials, $LiGaS_2$ [20, 21] and PtSeTe [27, 29] using density functional theory (DFT) and GW calculations. The calculated results of the band alignment, projected band structure, partial charge density and built-in electric field of $PtSeTe/LiGaS_2$ well support the idea of the D-scheme model. Finally, the $PtSeTe/LiGaS_2$ photocatalyst shows high photocatalytic performance, indicating the superiority of the proposed D-scheme model in photocatalytic water splitting. This dipole-scheme model is generalized and applicable to other polar materials, which can even be made of one polar and one non-polar material or one 2D and one 3D material. We seek for the experimental exploration along this path.

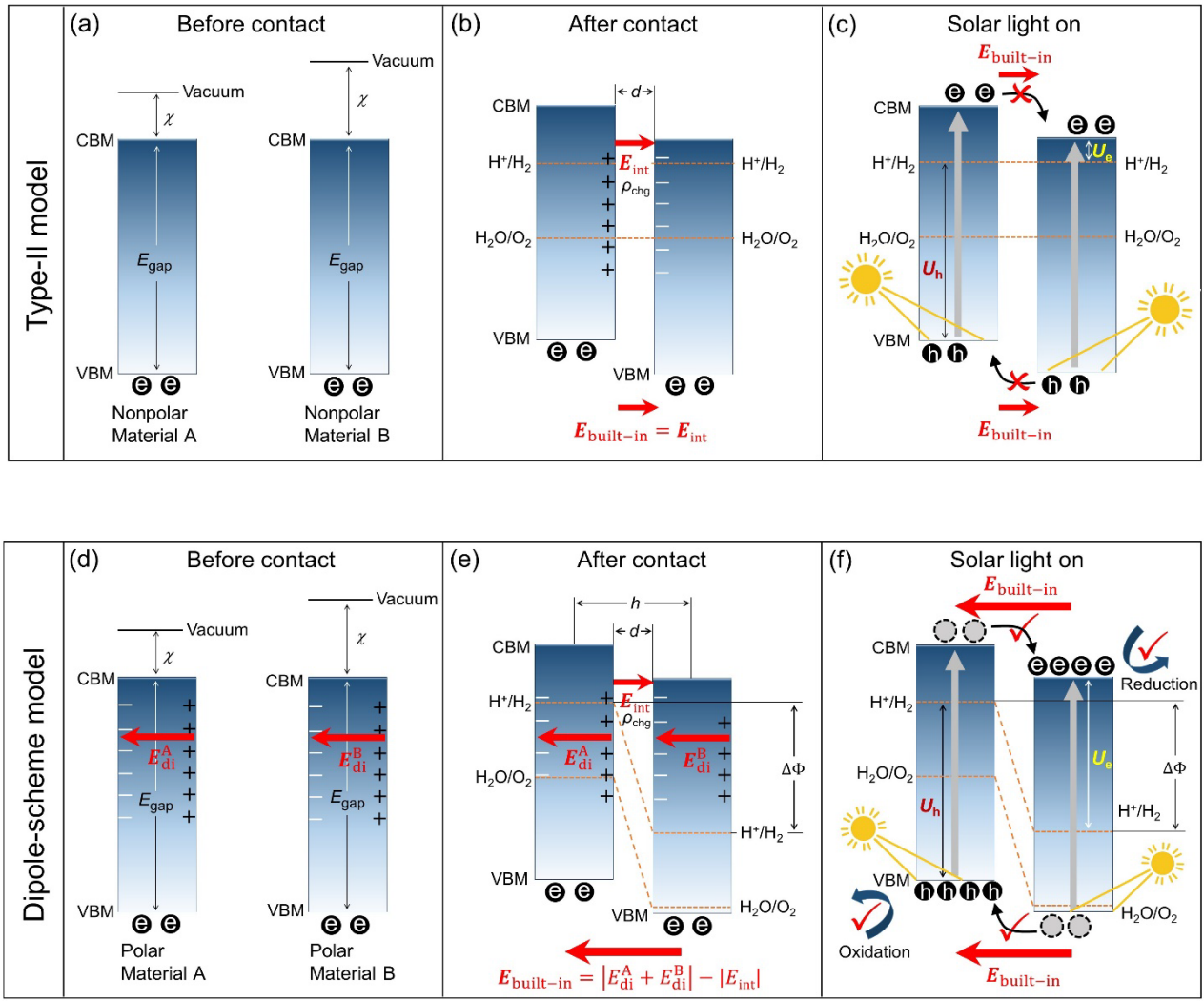

**Figure 1** Schematic of the charge-transfer mechanism in (a-c) conventional type-II and (d-f) dipole-scheme model before contact, after contact and under solar light. The grey arrows in (c) and (f) represent the electrons are excited from VBM to CBM by the solar light. The type-II heterojunction is made of two non-polar materials A and B. The interfacial electric field in the type-II heterojunction prevents the targeted charge transfer. Moreover, the type-II staggered band alignment results in small driving photovoltages ($U_e$, $U_h$). The D-scheme heterojunction has a large polarization field which facilitates the desired charge transfer, and generates a large driving force for redox reactions. It is noted that the D-scheme heterojunction has electric dipoles, meanwhile the charge-transfer route within the junction likes the letter "D". This is where it gets its name from.

## RESULTS AND DISCUSSION

**Dipole-Scheme Model and a Single Polar Example.** **Figure 1** shows the qualitative comparison between the proposed D-scheme model and conventional type-II model. The D-scheme model is distinguished by the direction of the built-in electric field ($E_{built-in}^D$), which is opposite to that in conventional type-II heterojunctions (**Figure 1e** vs **Figure 1b**). When the electrons and holes are excited by solar light in materials A and B, the built-in electric field in the type-II heterojunction ($E^{II}_{built-in}$) actually hampers the transfer of electrons from low work-function A to high work-function B as well as holes from B to A (see in **Figure 1c**). However, the direction of $E^{D}_{built-in}$ is opposite to $E^{II}_{built-in}$ which facilitates the targeted charge transfer route, greatly boosting the effective photogenerated carrier separation. Moreover, the driving force of H$_2$ reduction ($U_e$) is considerably enhanced by the electrostatic potential difference ($\Delta\Phi$) of two polar materials (**Figure 1f** vs **Figure 1c**).

We next present a quantitative comparison of the built-in electric field between the D-scheme and type-II model, $E^{D}_{built-in}$ and $E^{II}_{built-in}$. Assuming the interfacial charge transfer from the low work-function material A to high work-function material B in the D-scheme and type-II model is the same $\rho^{D}_{chg} = \rho^{II}_{chg}$. Then they have the same interfacial electric field from A to B (defined as the positive sign):

$$E^{II}_{int} = E^{D}_{int} \qquad (1)$$

Then, the total built-in electric field in the type-II heterojunction $E^{II}_{built-in}$ is just $E^{II}_{int}$ (see **Figure 1b**). In the D-scheme heterojunction, each polar material has its own polarization field $E^{A}_{di}$ and $E^{B}_{di}$ in addition to their interfacial field, $E^{D}_{int}$ as shown in **Figure 1e**. It is worth noting that the two polar materials are stacked purposely with the same dipole direction for having the maximum polarization, which is technically achievable in the experiment. [14, 31, 32, 41-43] Let's first ignore the interfacial dipole, then the electric field is

$$E_0 = -\Delta\Phi_0/h, \qquad (2)$$

where the negative sigh means its direction from material B to A, $\Delta\Phi_0 = \Delta\Phi_A + \Delta\Phi_B$ is the sum of the electrostatic potential difference in two polar materials, and $h$ is the effective thickness of the heterojunction. When the interfacial dipole between $A$ and $B$, $\Delta\Phi_{int}$, is taken into consideration, the total $\Delta\Phi = \Delta\Phi_0 - \Delta\Phi_{int}$ and $E^{D}_{built-in} = -\Delta\Phi/h$. Then, the relationship between $E^{D}_{built-in}$ and $E^{D}_{int}$ can be expressed as follows: [44, 45]

$$E^{D}_{built-in}/E^{D}_{int} = -\Delta\Phi/(\Delta\Phi_0 - \Delta\Phi), \qquad (3)$$

where the negative sigh means the direction of the built-in polarization field is opposite to the interfacial electric field (see **Figure 1e**). Since $E^{D}_{int} = E^{II}_{int} = E^{II}_{built-in}$ in Eq. (1), the ratio of the magnitude between $E^{D}_{built-in}$ and $E^{II}_{built-in}$ is $\Delta\Phi/(\Delta\Phi_0 - \Delta\Phi)$.

Based on Equation (3), we can give a quantitative description to the magnitude of $E_{\text{built-in}}^{\text{D}}$ and $E_{\text{built-in}}^{\text{II}}$ between our proposed D-scheme and the conventional type-II heterojunction with real materials. Because $\Delta\Phi$ usually is larger than $\Delta\Phi_{\text{int}}$, we will show that $E_{\text{built-in}}^{\text{D}}$ in two types of D-scheme prototypes can be 3-10 times of $E_{\text{built-in}}^{\text{II}}$ in the following sections.

Next, we construct two heterojunctions: PtTe$_2$/InS (PtTe$_2$ and InS are both non-polar) and *lite-version* D-scheme PtTe$_2$/LiInS$_2$ (single polar - only one material, LiInS$_2$, is polar) as a simple example to quantitatively compare the magnitude of $E_{\text{built-in}}^{\text{II}}$ and $E_{\text{built-in}}^{\text{D}}$ (see the configurations in **Figure S1**). It is found that for PtTe$_2$/InS, 0.062 $e$ ($\rho_{\text{chg}}^{\text{II}}$) is transferred from PtTe$_2$ to InS by the Bader analysis and the optimized interfacial distance ($d_1$) is 2.87 Å. For PtTe$_2$/LiInS$_2$, $\rho_{\text{chg}}^{\text{D}}$ is 0.072 $e$, $d_2$ is 2.87 Å, and the interfacial thickness is 6.23 Å ($h$). The $\Delta\Phi$ of polar PtTe$_2$/LiInS$_2$ is calculated to be 0.81 eV and $\Delta\Phi_0$ is 1.10 eV. Therefore, $E_{\text{built-in}}^{\text{D}}$ is around 3 times higher than $E_{\text{built-in}}^{\text{II}}$ according to Equation (3). Meanwhile, $E_{\text{built-in}}^{\text{D}}$ and $E_{\text{built-in}}^{\text{II}}$ are in the opposite direction. It is worth noting that the ratio of $E_{\text{built-in}}^{\text{D}}$ and $E_{\text{built-in}}^{\text{II}}$ can be further enhanced by using two polar materials, such as double polar PtSeTe/LiGaS$_2$. We will discuss this *full-version* D-scheme heterojunction in the details in following sections to demonstrate its high photocatalytic performance, including the band alignment, optical absorption spectrum, solar-to-hydrogen (STH) efficiency, built-in electric field, exciton binding energy, and Gibbs free energy of the hydrogen (HER) and oxygen evolution reaction (OER).

**Geometric and Electronic Structures of a Double Polar D-Scheme Example.** Two polar 2D materials, PtSeTe and LiGaS$_2$, have been reported with similar lattice constants of 3.80 and 3.81 Å, respectively.[20, 21, 27, 29] We thus can easily construct a heterojunction of PtSeTe/LiGaS$_2$ avoiding the lattice-mismatch issue. There are four potential stacking configurations based on the asymmetric structure and dipole direction (see **Figure S2**). The configuration in **Figure 2** is used in this work because it has the maximal polarization difference, $\Delta\Phi$, of 2.10 eV (**Figure S3**) as well as the suitable band gap and band alignment for the overall-water-splitting reactions (**Figure S4**). Besides the negative binding energy (**Table S1**), we further evaluate the dynamical and thermal stability of the heterojunction PtSeTe/LiGaS$_2$ by performing the phonon spectrum and *ab initio* molecular dynamics simulation. The phonon dispersion shows no imaginary modes in the Brillouin zone (**Figure 2c**).

Moreover, there are no obvious structural deformations or broken bonds in PtSeTe/LiGaS$_2$ at 300 K for 10 ps (**Figure 2d**). These results show the good dynamical and thermal stability of the heterojunction PtSeTe/LiGaS$_2$.

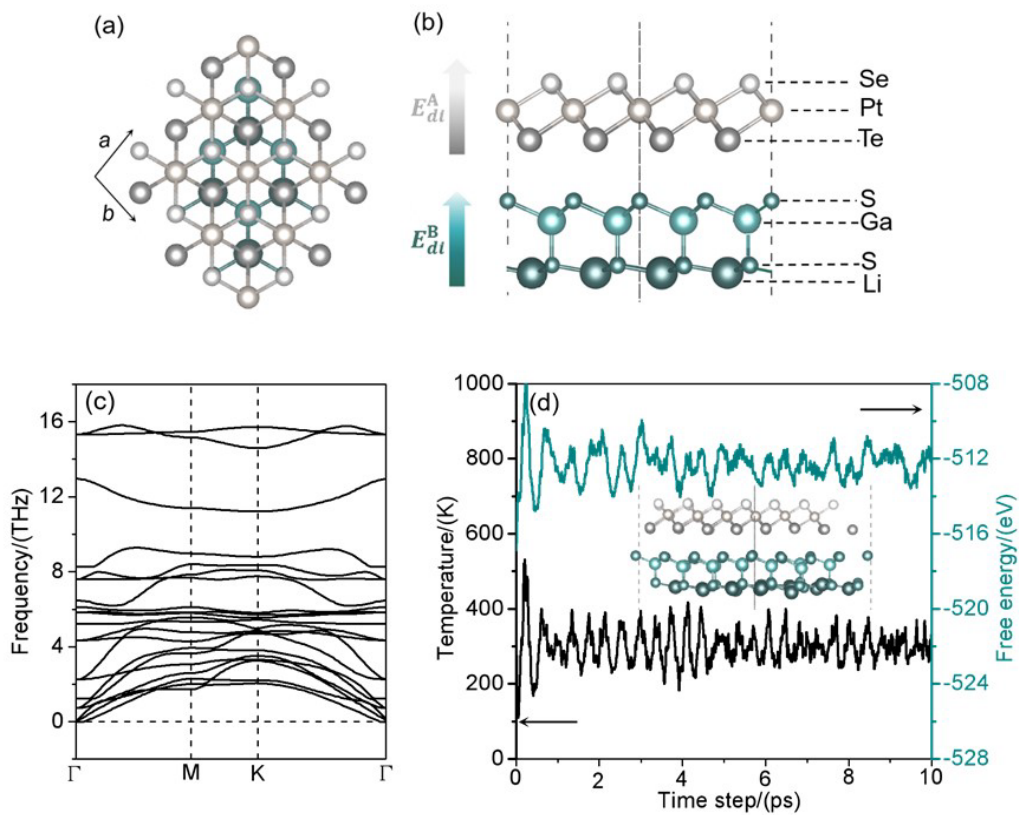

The HSE band structure of PtSeTe/LiGaS$_2$ is shown in **Figure S3a**. As can be seen, the heterojunction is semiconducting with a direct band gap of 1.10 eV at the HSE06 level. **Figures 3a** and **3b** show the projected band structure and partial charge density of the CBM and VBM. We can see that the state of CBM is mainly distributed on the LiGaS$_2$ layer, while the VBM is from the PtSeTe. Together with the direction of the built-in electric field, we know that the HER and OER occur on LiGaS$_2$ and PtSeTe with the driving photovoltages $U_e$ = 0.81 eV, $U_h$ = 2.39 eV, respectively (**Figure 3c**). It is known that to be an ideal overall-water splitting photocatalyst, it is thermodynamically required that the band edges span redox potentials of water splitting, *i.e.*, CBM above H$^+$/H$_2$ and VBM below H$_2$O/O$_2$. Based on the band alignment in **Figure 3c**, the D-scheme heterojunction PtSeTe/LiGaS$_2$ not only fulfills such criteria, but also has a large driving force of photovoltages, indicating its promising photocatalytic applications for overall water splitting.

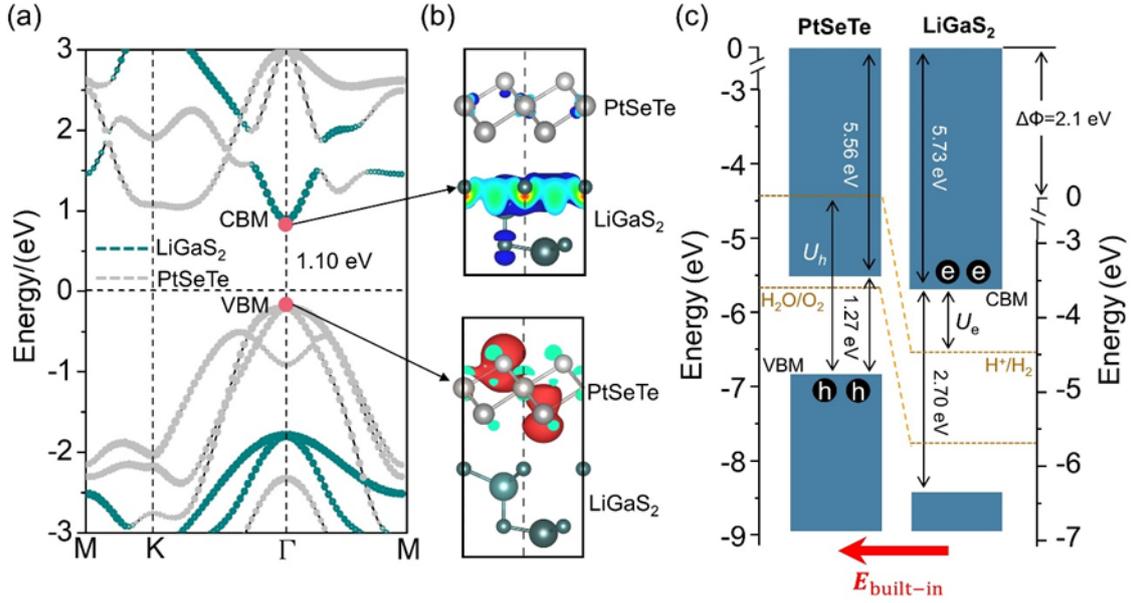

**Figure 3.** (a) The projected HSE band structure of the PtSeTe/LiGaS$_2$ heterojunction. The Fermi level is set to zero. (b) Partial charge densities of CBM and VBM. The isosurface value is 0.004 e/Å$^3$. (c) The band alignment with respect to the vacuum level (0 eV) of PtSeTe/LiGaS$_2$. The horizontal orange dashed lines represent reduction potential and oxidation potential of water splitting, respectively, which are bended by the polarization difference $\Delta\Phi$.

**Optical Adsorption and Conversion Efficiency.** After obtaining the band features of PtSeTe/LiGaS$_2$, we then investigate its light absorption spectrum (see **Figure 4**) by solving $G_0W_0$ + BSE approach using the following equation: [46]

$$\alpha(\omega) = \frac{\sqrt{2}\omega}{c}\left(\sqrt{\varepsilon_1^2 + \varepsilon_2^2} - \varepsilon_1\right)^{1/2} \quad (4)$$

where $c$ is the speed of light, $\varepsilon_1$ and $\varepsilon_2$ are the real part and imaginary part of the dielectric function, respectively. Our calculation results show that PtSeTe/LiGaS$_2$ exhibits significant absorbance of the visible and near infrared (NIR) spectrum, indicating its good light harvest capacity with a wide spectral responsibility. By compared to isolated LiGaS$_2$ [20, 21] and PtSeTe, [29] the absorption spectrum of PtSeTe/LiGaS$_2$ exhibits an obvious "red shift" and higher absorption coefficient of $10^5$ cm$^{-1}$. The good light-harvesting ability illustrates that the D-scheme PtSeTe/LiGaS$_2$ is a promising photocatalyst for water splitting. To evaluate the solar-to-hydrogen efficiency, we calculate four kinds of energy conversion efficiencies (see details in **SI**), *i.e.*, light absorption efficiency ($\eta_{abs}$), carrier utilization efficiency ($\eta_{cu}$), STH efficiency ($\eta_{STH}$) and corrected STH efficiency ($\eta'_{STH}$) of PtSeTe/LiGaS$_2$ (**Table S3**). We find that $\eta_{abs}$ of PtSeTe/LiGaS$_2$ reaches up to 81% and $\eta_{cu}$ value is higher than 60%. The $\eta'_{STH}$ is around 28% due to the small band gap, which is higher than that of monolayer M$_2$X$_3$,[8] and comparable to that of bilayer Li-III-VI compounds,[20] and B$_2$P$_6$.[37] Herein, the good light absorption and high utilization efficiency of PtSeTe/LiGaS$_2$ makes it economically feasible.

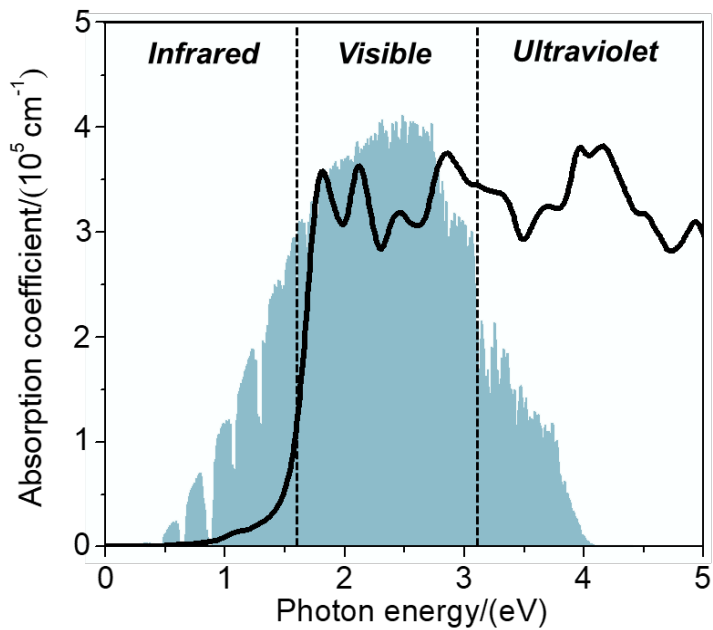

**Figure 4.** Optical absorption spectrum of heterojunction PtSeTe/LiGaS$_2$ calculated by $G_0W_0$ + BSE approach. It exhibits IR absorbance and a high absorption coefficient to $10^5$ cm$^{-1}$.

**Exciton-Dissociation and Carrier-Separation Ability.** As aforementioned, enough carriers involved in the redox reactions is critical for high solar-to-fuel efficiency as well as high photocatalytic performance. The photogenerated electron-hole pairs (excitons) should be effectively separated and transferred to the reduction and oxidation material of the heterojunction. We, next, investigate both the carrier-dissociation and carrier-separation ability in the D-scheme PtSeTe/LiGaS$_2$ heterojunction through the exciton binding energy ($E_{eb}$) and the built-in electric field, respectively. As we know a small exciton binding energy is desired for efficient carrier dissociation, which is defined by $E_{eb} = E_{QP} - E_{opt}$, where $E_{QP}$ and $E_{opt}$ are direct quasiparticle $G_0W_0$ band gap and the first peak of dielectric imaginary part using $G_0W_0$ + BSE approach, respectively. The calculated exciton binding energy of PtSeTe/LiGaS$_2$ is only 0.24 eV (**Figure S5c**), less than that of isolated LiGaS$_2$ (0.77 eV) [20, 21] and PtSeTe (0.33 eV), [29] and much smaller than well-studied 2D photocatalysts, such as g-C$_3$N$_4$

(1.20 eV) [47] and MoSSe (0.85 eV). [48]

To evaluate the polarization field in PtSeTe/LiGaS$_2$, we calculate $\Delta\Phi$ (2.10 eV **Figure S3**) of the double polar PtSeTe/LiGaS$_2$, and then obtain $E_{built-in}^{D}$ is 0.33 V/Å which is around about 10 times stronger than the interfacial field. Moreover, $\Delta\Phi$ is larger than the valence band offset (1.60 eV) and conductance band offset (0.17 eV), which means that the polarization field is strong enough to drive the inter-transfer of photogenerated carriers between PtSeTe and LiGaS$_2$. Such large built-in polarization field can effectively prevent the recombination of the free electrons and holes, [19, 38, 45] and push them into LiGaS$_2$ and PtSeTe to participate in the reduction and oxidation reaction, respectively.

**OER and HER of Overall Water Splitting.** Finally, after accumulating a large amount of free electrons and holes in reduction LiGaS$_2$ and oxidation PtSeTe, respectively, we study the redox reaction efficiency of D-scheme PtSeTe/LiGaS$_2$ by calculating the Gibbs free energy of OER and HER (see detailed calculations in **SI**). The PtSeTe/LiGaS$_2$ with the most stable adsorptions of intermediates (OH*, O*, OOH* and H*) during the redox reaction are shown in **Figures 5a** and **5b**, where the OER and HER process occurs on the PtSeTe and LiGaS$_2$ layer, respectively. **Figure 5c** depicts the free-energy profiles of the four-step OER. Here $\Delta i$ represents the free energy change in step $i$. In the dark environment at pH = 0, we find that $\Delta_1$, $\Delta_2$ and $\Delta_3$ are positive, and the largest overpotential is 2.37 eV in the 3$^{rd}$ step ($\Delta_3$). Such a large overpotential suggests the oxidation reaction of water splitting cannot proceed smoothly. However, under light irradiation, the photoexcited holes supply an external potential $U_h$ of 2.39 eV, all steps in the half oxidation reaction of water splitting are downhill, indicating a spontaneous OER reaction in PtSeTe/LiGaS$_2$ under solar light. For the HER part, the free energy change for step 1 ($\Delta G_H$) is of 1.11 eV (**Figure 5d**) without light irradiation. After switching on the light, the photoexcited electrons can supply a large potential $U_e$ of 0.81 eV, resulting in a small $\Delta G_H$ of 0.30 eV. This $\Delta G_H$ is much lower than that of reported photocatalysts of monolayer MoSi$_2$N$_4$ (1.51 eV) and MoSiGeN$_4$ (1.76 eV) [34] and PdSSe (1.55 eV). [49] The OER and HER results indicate that the enhanced photocatalytic activity and performance of D-scheme PtSeTe/LiGaS$_2$, which is favorable to its high overall photocatalytic efficiency.

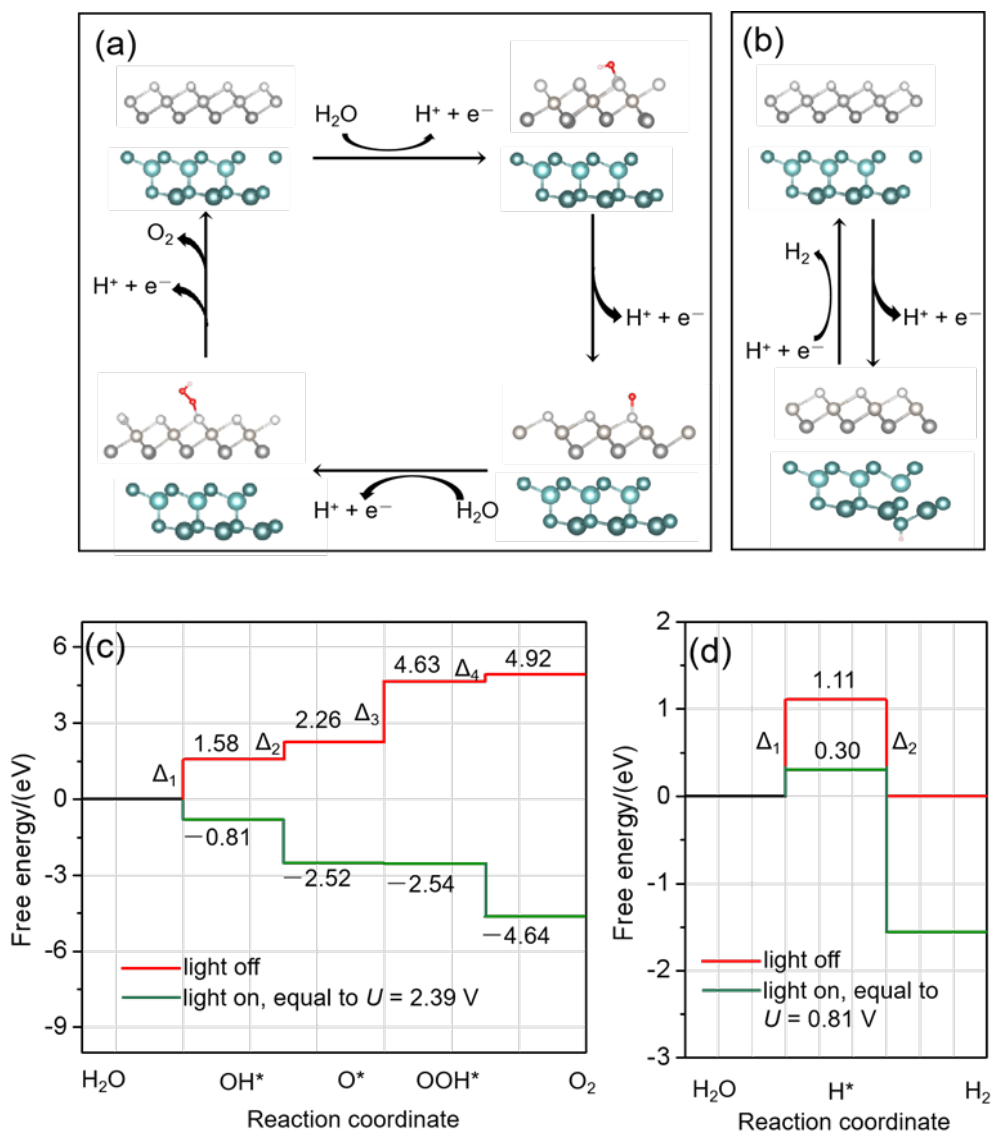

**Figure 5.** Proposed photocatalytic pathways of OER (a) and HER (b) with the most energetically favorable absorbed intermediates (OH*, O*, OOH* and H*) on the D-scheme PtSeTe/LiGaS$_2$. Gibbs free energy profiles of the OER (a) and HER (b) on the HBL PtSeTe/LiGaS$_2$ at pH = 0.

## CONCLUSIONS

In summary, we develop a new model for photocatalytic overall water splitting, in which a D-scheme heterojunction with two different polar materials is constructed. Note that the D-scheme heterojunctions exclude the 2D bilayer polar material which is still a single photocatalyst and each monolayer has the same work function and band gap, suffering the issues of common single photocatalysts. Different from previously reported type-II, Z-scheme and S-scheme models, [7] the D-scheme heterojunctions have a narrow gap (may less than the water redox gap of 1.23 eV), large polarization field (opposite to the interfacial field), and high photovoltages for OER and HER. Moreover, all photogenerated electrons and holes are preserved to engage in photocatalytic reactions. These advantages act as the driving forces for the effective absorption of solar light, separation of carriers, high solar-to-fuel efficiency, and high redox ability. This D-scheme model is generalized and applicable for heterojunctions with either single or double polar materials. Taking PtSeTe/LiGaS$_2$ (a heterojunction with two 2D polar materials) as an example, we show that its band gap is only 1.10 eV (near-infrared adsorption), but it has large redox driving forces ($U_h = 2.39$ eV, $U_e = 0.81$ eV) and then a high efficiency of OER and HER. The built-in polarization field by dipole moments of polar PtSeTe and LiGaS$_2$ can provide a high driving photo-energy of 2.1 eV for targeted both electron and hole transfer between the reduction and oxidation materials. Above all, this work provides not only a novel model, but also materials realization of high-performance photocatalysts for overall water splitting utilizing the emerging 2D polar materials.

## COMPUTATIONAL METHODS

All the calculations were performed within the Vienna Ab initio Simulation Package (VASP). [46] The generalized gradient approximation (GGA) of Perdew, Burke and Ernzerhof (PBE) and the projected augmented wave (PAW) method [50] were used to describe the exchange correlation potential and the ion-electron interaction. Due to a vertical dipole moment in the polar systems, the dipole correction was introduced along the *z*-direction. The energy cutoff of plane wave was set to 500 eV and Γ-centered k-point mesh of 18 × 18 × 1 were used to sample the Brillouin zone. A vacuum layer of thickness 20 Å was included between neighboring supercells to avoid interaction between them. The DFT-D2 correction of Grimme [51] was adopted to better describe the van der Waals interactions between the adjacent images. The convergence criteria for the total energy and Hellmann−Feynman force were set to 1×10$^{-5}$ eV and 1×10$^{-2}$ eV/Å, respectively. Spin orbit coupling (SOC) was also considered in the self-consistent calculations. The details of phonon dispersion, molecular dynamics simulation, HSE06 and GW calculations are available in the Supporting Information.

## ACKNOWLEDEMENTS

The authors thank X. Q. Liu for his helpful discussion. X. Gao acknowledges the financial support from the China Scholarship Council (No.202006120280). X. Gao, Y. Q. Shen, J. J. Liu, L. L. Lv, M. Zhou and Z. X. Zhou gratefully acknowledge the support from the National Natural Science Foundation of China (No.11204053 and No.11074059). This research is also partially supported by the Ministry of Education, Singapore, under its MOE Tier 1 Grant (R-265-000-691-114) and MOE AcRF Tier 2 Award MOE2019-T2-2-30. Authors acknowledge the use of computational resources at the National Supercomputing Centre (NSCC) of Singapore and the High Performance Computing at the National University of Singapore.

‡ Contributed equally to this work

*Corresponding author: Yanqing Shen, E-mail: shenyanqing2004@163.com.

Lei Shen, E-mail: shenlei@nus.edu.sg.

## REFERENCES

[1] Moniz, S. J. A.; Shevlin, S. A.; Martin, D. J.; Guo, Z.-X.; Tang, J. Visible-Light Driven Heterojunction Photocatalysts for Water Splitting – A Critical Review. *Energ. Environ. Sci.* **2015,** *8*, 731-759.

[2] Stolarczyk, J. K.; Bhattacharyya, S.; Polavarapu, L.; Feldmann, J. Challenges and Prospects in Solar Water Splitting and $CO_2$ Reduction with Inorganic and Hybrid Nanostructures. *ACS Cataly.* **2018,** *8*, 3602-3635.

[3] Li, X.; Li, Z.; Yang, J. Proposed Photosynthesis Method for Producing Hydrogen from Dissociated Water Molecules Using Incident Near-Infrared Light. *Phys. Rev. Lett.* **2014,** *112*, 018301.

[4] Takanabe, K. Photocatalytic Water Splitting: Quantitative Approaches toward Photocatalyst by Design. *ACS Catal.* **2017,** *7*, 8006-8022.

[5] Gao, C.; Wang, J.; Xu, H.; Xiong, Y. Coordination Chemistry in the Design of Heterogeneous Photocatalysts. *Chem. Soc. Rev.* **2017,** *46*, 2799-2823.

[6] Fu, C. F.; Wu, X.; Yang, J. Material Design for Photocatalytic Water Splitting from a Theoretical Perspective. *Adv. Mater.* **2018**, *30*, e1802106.

[7] Xu, Q.; Zhang, L.; Cheng, B.; Fan, J.; Yu, J. S-Scheme Heterojunction Photocatalyst. *Chem* **2020**, *6*, 1543-1559.

[8] Fu, C. F.; Sun, J.; Luo, Q.; Li, X.; Hu, W.; Yang, J. Intrinsic Electric Fields in Two-dimensional Materials Boost the Solar-to-Hydrogen Efficiency for Photocatalytic Water Splitting. *Nano Lett.* **2018**, *18*, 6312-6317.

[9] Zhao, P.; Ma, Y.; Lv, X.; Li, M.; Huang, B.; Dai, Y. Two-Dimensional $III_2$-$VI_3$ Materials: Promising Photocatalysts for Overall Water Splitting under Infrared Light Spectrum. *Nano Energy* **2018**, *51*, 533-538.

[10] Liao, Y.; Zhang, Z.; Gao, Z.; Qian, Q.; Hua, M. Tunable Properties of Novel $Ga_2O_3$ Monolayer for Electronic and Optoelectronic Applications. *ACS Appl. Mater. Interfaces* **2020,** *12*, 30659-30669.

[11] Wang, P.; Liu, H.; Zong, Y.; Wen, H.; Xia, J. B.; Wu, H. B. Two-Dimensional $In_2X_2X'$ (X and X' = S, Se, and Te) Monolayers with an Intrinsic Electric Field for High-Performance Photocatalytic and Piezoelectric Applications. *ACS Appl. Mater. Interfaces* **2021**, *13*, 34178-34187.

[12] Ding, W.; Zhu, J.; Wang, Z.; Gao, Y.; Xiao, D.; Gu, Y.; Zhang, Z.; Zhu, W. Prediction of Intrinsic Two-Dimensional Ferroelectrics in $In_2Se_3$ and other $III_2$-$VI_3$ van der Waals Materials. *Nat. Commun.* **2017,** *8*, 14956.

[13] Zhou, J.; Zeng, Q.; Lv, D.; Sun, L.; Niu, L.; Fu, W.; Liu, F.; Shen, Z.; Jin, C.; Liu, Z. Controlled Synthesis of High-Quality Monolayered alpha-$In_2Se_3$ via Physical Vapor Deposition. *Nano Lett.* **2015,** *15*, 6400-5.

[14] Almeida, G.; Dogan, S.; Bertoni, G.; Giannini, C.; Gaspari, R.; Perissinotto, S.; Krahne, R.; Ghosh, S.; Manna, L. Colloidal Monolayer beta-$In_2Se_3$ Nanosheets with High Photoresponsivity. *J. Am. Chem. Soc.* **2017,** *139*, 3005-3011.

[15] Zhou, Y.; Wu, D.; Zhu, Y.; Cho, Y.; He, Q.; Yang, X.; Herrera, K.; Chu, Z.; Han, Y.; Downer, M. C.; Peng, H.; Lai, K. Out-of-Plane Piezoelectricity and Ferroelectricity in Layered alpha-$In_2Se_3$ Nanoflakes. *Nano Lett.* **2017,** *17*, 5508-5513.

[16] Xiao, J.; Zhu, H.; Wang, Y.; Feng, W.; Hu, Y.; Dasgupta, A.; Han, Y.; Wang, Y.; Muller, D. A.; Martin, L. W.; Hu, P.; Zhang, X. Intrinsic Two-Dimensional Ferroelectricity with Dipole Locking. *Phys. Rev. Lett.* **2018,** *120*, 227601.

[17] Li, S.; Shi, M.; Yu, J.; Li, S.; Lei, S.; Lin, L.; Wang, J. Two-Dimensional Blue-Phase CX (X = S, Se) Mmonolayers with High Carrier Mobility and Tunable Photocatalytic Water Splitting Capability. *Chinese Chem. Lett.* **2021,** *32*, 1977-1982.

[18] Ji, Y.; Yang, M.; Dong, H.; Hou, T.; Wang, L.; Li, Y. Two-Dimensional Germanium Monochalcogenide Photocatalyst for Water Splitting under Ultraviolet, Visible to Near-Infrared Light. *Nanoscale* **2017,** *9*, 8608-8615.

[19] Yang, H.; Ma, Y.; Zhang, S.; Jin, H.; Huang, B.; Dai, Y. GeSe@SnS: Stacked Janus Structures for Overall Water Splitting. *J. Mater. Chem. A* **2019,** *7*, 12060-12067.

[20] Fan, Y.; Song, X.; Qi, S.; Ma, X.; Zhao, M. Li-III-VI Bilayers for Efficient Photocatalytic Overall Water Splitting: The Role of Intrinsic Electric Field. *J. Mater. Chem. A* **2019,** *7*, 26123-26130.

[21] Xu, W.; Wang, R.; Zheng, B.; Wu, X.; Xu, H. Two-Dimensional Li-Based Ternary Chalcogenides for Photocatalysis. *J. Phys. Chem. Lett.* **2019,** *10*, 6061-6066.

[22] Vu, T. V.; Nguyen, C. V.; Phuc, H. V.; Lavrentyev, A. A.; Khyzhun, O. Y.; Hieu, N. V.; Obeid, M. M.; Rai, D. P.; Tong, H. D.; Hieu, N. N. Theoretical Prediction of Electronic, Transport, Optical, and Thermoelectric Properties of Janus Monolayers $In_2XO$ (X = S, Se, Te). *Phys. Rev. B* **2021,** *103*, 085422.

[23] Xia, C.; Xiong, W.; Du, J.; Wang, T.; Peng, Y.; Li, J. Universality of Electronic Characteristics and Photocatalyst Applications in the Two-Dimensional Janus Transition Metal Dichalcogenides. *Phys. Rev. B* **2018,** *98*, 165424.

[24] Ju, L.; Bie, M.; Tang, X.; Shang, J.; Kou, L. Janus WSSe Monolayer: An Excellent Photocatalyst for Overall Water Splitting. *ACS Appl. Mater. Interfaces* **2020,** *12*, 29335-29343.

[25] Zhao, P.; Liang, Y.; Ma, Y.; Huang, B.; Dai, Y. Janus Chromium Dichalcogenide Monolayers with Low Carrier Recombination for Photocatalytic Overall Water-Splitting under Infrared Light. *J. Phys. Chem. C* **2019,** *123*, 4186-4192.

[26] Varjovi, M. J.; Yagmurcukardes, M.; Peeters, F. M.; Durgun, E. Janus Two-Dimensional Transition Metal Dichalcogenide Oxides: First-Principles Investigation of WXO Monolayers with X = S, Se, and Te. *Phys. Rev. B* **2021,** *103*, 195438.

[27] Ersan, F.; Ataca, C. Janus $PtX_nY_{2-n}$ (X, Y = S, Se, Te; $0 \leq n \leq 2$) Monolayers for Enhanced Photocatalytic Water Splitting. *Phys. Rev. Appl.* **2020,** *13*, 064008.

[28] Peng, R.; Ma, Y.; Huang, B.; Dai, Y. Two-Dimensional Janus PtSSe for Photocatalytic Water Splitting under the Visible or Infrared Light. *J. Mater. Chem. A* **2019,** *7*, 603-610.

[29] Gao, X.; Shen, Y.; Liu, J.; Lv, L.; Zhou, M.; Zhou, Z.; Feng, Y. P.; Shen, L. Boosting the Photon Absorption, Exciton Dissociation, and Photocatalytic Hydrogen- and Oxygen-Evolution Reactions by Built-in electric Fields in Janus Platinum Dichalcogenides. *J. Mater. Chem. C* **2021,** *9*, 15026-15033.

[30] Shen, H.; Zhang, Y.; Wang, G.; Ji, W.; Xue, X.; Zhang, W. Janus PtXO (X = S, Se) Monolayers: the Visible Light Driven Water Splitting Photocatalysts with High Carrier Mobilities. *Phys. Chem. Chem. Phys.* **2021,** *23*, 21825-21832.

[31] Lu, A. Y.; Zhu, H.; Xiao, J.; Chuu, C. P.; Han, Y.; Chiu, M. H.; Cheng, C. C.; Yang, C. W.; Wei, K. H.; Yang, Y.; Wang, Y.; Sokaras, D.; Nordlund, D.; Yang, P.; Muller, D. A.; Chou, M. Y.; Zhang, X.; Li, L. J. Janus Monolayers of Transition Metal Dichalcogenides. *Nat. Nanotechnol.* **2017,** *12*, 744-749.

[32] Zhang, J.; Jia, S.; Kholmanov, I.; Dong, L.; Er, D.; Chen, W.; Guo, H.; Jin, Z.; Shenoy, V. B.; Shi, L.; Lou, J. Janus Monolayer Transition-Metal Dichalcogenides. *ACS Nano* **2017,** *11*, 8192-8198.

[33] Sant, R.; Gay, M.; Marty, A.; Lisi, S.; Harrabi, R.; Vergnaud, C.; Dau, M. T.; Weng, X.; Coraux, J.; Gauthier, N.; Renault, O.; Renaud, G.; Jamet, M. Synthesis of Epitaxial Monolayer Janus SPtSe. *npj 2D Mater. and Appl.* **2020,** *4*, 41.

[34] Yu, Y.; Zhou, J.; Guo, Z.; Sun, Z. Novel Two-Dimensional Janus $MoSiGeN_4$ and $WSiGeN_4$ as Highly Efficient Photocatalysts for Spontaneous Overall Water Splitting. *ACS Appl. Mater. Interfaces* **2021,** *13*, 28090-28097.

[35] Zhang, Y.; Sa, B.; Miao, N.; Zhou, J.; Sun, Z. Computational Mining of Janus $Sc_2C$-Based MXenes for Spintronic, Photocatalytic, and Solar Cell Applications. *J. Mater. Chem. A* **2021,** *9*, 10882-10892.

[36] Qi, S.; Fan, Y.; Wang, J.; Song, X.; Li, W.; Zhao, M. Metal-Free Highly Efficient Photocatalysts for Overall Water Splitting: $C_3N_5$ Multilayers. *Nanoscale* **2020,** *12*, 306-315.

[37] Sun, M.; Schwingenschlögl, U. $B_2P_6$: A Two-Dimensional Anisotropic Janus Material with Potential in Photocatalytic Water Splitting and Metal-Ion Batteries. *Chem. Mater.* **2020,** *32*, 4795-4800.

[38] Fan, Y.; Song, X.; Ai, H.; Li, W.; Zhao, M. Highly Efficient Photocatalytic $CO_2$ Reduction in Two-Dimensional Ferroelectric $CuInP_2S_6$ Bilayers. *ACS Appl. Mater. Interfaces* **2021,** *13*, 34486-34494.

[39] Ju, L.; Shang, J.; Tang, X.; Kou, L. Tunable Photocatalytic Water Splitting by the Ferroelectric Switch in a 2D $AgBiP_2Se_6$ Monolayer. *J. Am. Chem. Soc.* **2020,** *142*, 1492-1500.

[40] Ju, L.; Tan, X.; Mao, X.; Gu, Y.; Smith, S.; Du, A.; Chen, Z.; Chen, C.; Kou, L. Controllable $CO_2$ Electrocatalytic Rduction via Ferroelectric Switching on Single Atom Anchored $In_2Se_3$ Monolayer. *Nat. Commun.* **2021,** *12*, 1-10.

[41] Chaves, A.; Azadani, J. G.; Alsalman, H.; da Costa, D. R.; Frisenda, R.; Chaves, A. J.; Song, S. H.; Kim, Y. D.; He, D.; Zhou, J.; Castellanos-Gomez, A.; Peeters, F. M.; Liu, Z.; Hinkle, C. L.; Oh, S.-H.; Ye, P. D.; Koester, S. J.; Lee, Y. H.; Avouris, P.; Wang, X.; Low, T. Bandgap Engineering of Two-Dimensional Semiconductor Materials. *npj 2D Mater. and Appl.* **2020,** *4*, 29.

[42] Wang, X.; Zhu, C.; Deng, Y.; Duan, R.; Chen, J.; Zeng, Q.; Zhou, J.; Fu, Q.; You, L.; Liu, S.; Edgar, J. H.; Yu, P.; Liu, Z. Van der Waals Engineering of Ferroelectric Heterostructures for Long-Retention Memory. *Nat. Commun.* **2021,** *12*, 1109.

[43] Lin, B.; Chaturvedi, A.; Di, J.; You, L.; Lai, C.; Duan, R.; Zhou, J.; Xu, B.; Chen, Z.; Song, P.; Peng, J.; Ma, B.; Liu, H.; Meng, P.; Yang, G.; Zhang, H.; Liu, Z.; Liu, F. Ferroelectric-Field Accelerated Charge Transfer in 2D $CuInP_2S_6$ Heterostructure for Enhanced Photocatalytic $H_2$ Evolution. *Nano Energy* **2020,** *76*, 104972.

[44] Gava, P.; Lazzeri, M.; Saitta, A. M.; Mauri, F. Ab Initio Study of Gap Opening and Screening Effects in Gated Bilayer Graphene. *Phys. Rev. B* **2009,** *79*, 165431.

[45] Niu, X.; Bai, X.; Zhou, Z.; Wang, J. Rational Design and Characterization of Direct Z-Scheme Photocatalyst for Overall Water Splitting from Excited State Dynamics Simulations. *ACS Catal.* **2020,** *10*, 1976-1983.

[46] Adolph, B.; Furthmüller, J.; Bechstedt, F. Optical Properties of Semiconductors Using Projector-Augmented Waves. *Phys. Rev. B* **2001,** *63*, 125108.

[47] Wang, W. L.; Kaxiras, E. Efficient Calculation of the Effective Single-Particle Potential and its Application in Electron Microscopy. *Phys. Rev. B* **2013,** *87*, 085103.

[48] Li, F.; Wei, W.; Huang, B.; Dai, Y. Excited-State Properties of Janus Transition-Metal Dichalcogenides. *J. Phys. Chem. C* **2019,** *124*, 1667-1673.

[49] Xiao, F.; Lei, W.; Wang, W.; Xu, L.; Zhang, S.; Ming, X. Pentagonal Two-Dimensional Noble-Metal Dichalcogenide PdSSe for Photocatalytic Water Splitting with Pronounced Optical Absorption and Ultrahigh Anisotropic Carrier Mobility. *J. Mater. Chem. C* **2021,** *9*, 7753-7764.

[50] Grimme, S. Semiempirical GGA-Type Density Functional Constructed with a Long-Range Dispersion Correction. *J. Comput. Chem.* **2006,** *27*, 1787-99.

[51] Grimme, S. Accurate Description of van der Waals Complexes by Density Functional Theory Including Empirical Corrections. *J. Comput. Chem.* **2004,** *25*, 1463-73.